\documentclass[apj, 8pt]{emulateapj}
\usepackage{apjfonts}
\usepackage{epsf}

\usepackage[dvipdfm]{hyperref}
\hypersetup{
	bookmarks=true,
	bookmarksnumbered=true,
	bookmarksopen=true,
	pdftitle = {The First Systematic Survey for Lyman Alpha Emitters at z=7.3 with Red-sensitive Subaru/Suprime--Cam},
	pdfauthor = {Takatoshi Shibuya},
	pdfkeywords={cosmology: observations --- early universe --- galaxies: formation --- galaxies: high-redshift}
	pdfpagemode=UseThumbs
}

\renewcommand{\thefootnote}{\alph{footnote}}

\slugcomment{Accepted for Publication in The Astrophysical Journal}

\shorttitle{Survey for LAEs at $z=7.3$}
\shortauthors{T.Shibuya, N.Kashikawa, K.Ota, M.Iye, M.Ouchi, H.Furusawa, K.Shimasaku, and T.Hattori}

\begin{document}

\title{The First Systematic Survey for Lyman Alpha Emitters at $z=7.3$ \\with 
Red-sensitive Subaru/Suprime--Cam\altaffilmark{\ddag}}


\author{Takatoshi SHIBUYA\altaffilmark{1,2}, Nobunari KASHIKAWA\altaffilmark{1,2}, Kazuaki OTA\altaffilmark{3}, 
Masanori IYE\altaffilmark{1,2,4},  Masami OUCHI\altaffilmark{5,6}, \\Hisanori FURUSAWA\altaffilmark{2}, Kazuhiro SHIMASAKU\altaffilmark{7,8},  
and Takashi HATTORI\altaffilmark{9}}
\email{takatoshi.shibuya_at_nao.ac.jp}

\altaffiltext{1}{Department of Astronomical Science, The Graduate University for Advanced Studies, Mitaka, Tokyo, 181-8588, Japan.}
\altaffiltext{2}{Optical and Infrared Astronomy Division, National Astronomical Observatory, Mitaka, Tokyo 181-8588, Japan.}
\altaffiltext{3}{Department of Astronomy, Kyoto University, Kitashirakawa-Oiwake-cho, Sakyo-ku, Kyoto 606-8502, Japan.}
\altaffiltext{4}{Department of Astronomy, The University of Tokyo, Mitaka, Tokyo 181-8588, Japan.}
\altaffiltext{5}{Institute for Cosmic Ray Research, University of Tokyo, Kashiwa 277-8582, Japan.}
\altaffiltext{6}{Institute for the Physics and Mathematics of the Universe (IPMU), TODIAS, University of Tokyo, 5-1-5 Kashiwanoha, Kashiwa, Chiba 277-8583, Japan.}
\altaffiltext{7}{Department of Astronomy, Graduate School of Science, The University of Tokyo, Tokyo 113-0033, Japan.}
\altaffiltext{8}{Research Center for the Early Universe, Graduate School of Science, The University of Tokyo, Tokyo 113-0033, Japan.}
\altaffiltext{9}{Subaru Telescope, 650 North A'ohoku Place, Hilo, HI 96720.}
\altaffiltext{\ddag}{Based on data obtained with the Subaru Telescope and the W. M. Keck Observatory. 
The Subaru Telescope is operated by the National Astronomical Observatory of Japan. 
The W. M. Keck Observatory is operated as a scientific partnership among
the California Institute of Technology, the University of California,
and the National Aeronautics and Space Administration.}


\begin{abstract}
We have performed deep imaging surveys for Lyman $\alpha$ emitters (LAEs) at redshift $\sim 7.3$ in 
two blank fields, the Subaru Deep Field (SDF) and the Subaru/XMM-Newton Deep survey Field (SXDF), 
using the Subaru/Suprime-Cam equipped  with new red-sensitive CCDs and a new narrow-band filter, 
NB1006 ($\lambda_c=10052$\AA, FWHM $\Delta \lambda=214$\AA).
We identified four objects as LAE candidates that exhibit luminosity excess in NB1006. 
By carrying out deep follow-up spectroscopy for three of them using Subaru/FOCAS and Keck/DEIMOS, 
a definitively asymmetric emission line is detected for one of them, SXDF-NB1006-2.  Assuming this line is Ly$\alpha$,  
this object is a LAE at $z=7.215$ which has luminosity of $1.2^{+1.5}_{-0.6} \times 10^{43} $ erg s$^{-1}$ 
and a weighted skewness S$_\omega=4.90\pm 0.86$. 
Another object, SDF-NB1006-2, shows variable photometry and is thus probably a quasar (QSO) or an active galactic nucleus (AGN).  
It shows an asymmetric emission line at $10076$\AA\, which may be due to either Ly$\alpha$ at $z=7.288$ or $[$O {\sc ii}$]$ at $z=1.703$.
The third object, SDF-NB1006-1, is likely a galaxy with temporal luminosity enhancement associated with a supernova explosion, 
as the brightness of this object varies between the observed epochs. Its spectrum does not show any emission lines. 
The inferred decrease in the number density of LAEs toward higher redshift is
$n_{Ly\alpha}^{z=7.3}/n_{Ly\alpha}^{z=5.7} = 0.05^{+0.11}_{-0.05}$ from $z=5.7$ to $7.3$ 
down to $L^{Ly\alpha}=1.0 \times 10^{43}$ erg s$^{-1}$. 
The present result is consistent with the interpretation in previous studies that the neutral hydrogen fraction 
is rapidly increasing from $z=5.7$ to $7.3$. 
\end{abstract}

\keywords{cosmology: observations --- early universe --- galaxies: formation --- galaxies: high-redshift}


\section{INTRODUCTION}

Cosmic reionization is one of the most important events in the early history of the universe. 
It is the drastic phase transition of the intergalactic hydrogen medium (IGM) from neutral to ionized;
therefore it is closely related to the formation and evolution of the first stars and galaxies. 
How and when this cosmic reionization process took place remains an issue of great interest  
in the studies of the early history of the universe and many observational studies have been 
made to probe early galaxies during the last decade \citep{2011arXiv1110.0195I}. 
The latest measurements of the polarization of the cosmic microwave 
background (CMB) by the Wilkinson Microwave Anisotropy Probe (WMAP) constrained the optical depth of 
electron scattering during reionization, suggesting that the probable redshift of reionization, 
if it took place instantaneously, was $z=11.0 \pm 1.4$ \citep{2009ApJS..180..306D}. 
Measurements of Gunn-Peterson (GP) troughs (Gunn  \& Peterson 1965) on spectra of high-z quasars sampled by 
Sloan Digital Sky Survey (SDSS) imply that reionization had been completed by $z\sim 6$, when the fraction of IGM neutral hydrogen is 
estimated to be very small, $x^{z\sim 6.2}_{HI}\sim 0.01-0.04$ \citep{2006AJ....132..117F}, where
 $x_{HI}$ is defined as $x_{HI} = n_{HI}/n_{H}$ with $n_{HI}$ and $n_{H}$ being the number density 
of neutral hydrogen and all hydrogen in the universe, respectively. 

Another promising technique to probe the reionization process is to use Lyman $\alpha$ emitters (LAEs).
Narrow-band (NB)  (bandwidths of $\sim 100-200$\AA) imaging can efficiently isolate LAEs, which prominently 
emit Ly$\alpha$ from their interstellar gas ionized by massive stars. 
Comparing Ly$\alpha$ luminosity functions (LFs) of LAEs before and after cosmic reionization 
provides an approach to determine the epoch of end of reionization \citep{1999ApJ...518..138H, 2004ApJ...617L...5M}, 
because Lyman $\alpha$ photons are sensitive to neutral IGM abundances.
In the reionization era,  Lyman $\alpha$ photons would be scattered at higher redshifts 
due to the damping wing from the surrounding neutral IGM. 
Therefore, the observed abundance of LAEs at this epoch should vary with the neutral fraction of IGM hydrogen. 
Using LAEs to probe reionization is sensitive to neutral IGM fractions $x_{HI}>10^{-3}$. 
This is an advantage over the GP test which is not sensitive at $x_{HI}>10^{-3}$. 
LAE imaging surveys are a good observational method to probe the epoch 
of the cosmic reionization, although it is difficult to distinguish the effects of reionization from those of 
galaxy evolution.


Due to their very low number density, wide field deep survey observations are essential to construct a statistical 
sample of high-redshift LAEs. Wide field coverage also allows us to overcome the uncertainties of 
cosmic variance, as well as potentially large cosmic variance. 
The Subaru Prime Focus Camera \citep[Suprime-Cam]{2002PASJ...54..833M} has ten 
MIT/LL $2048$ pix $\times4096$ pix CCDs, located at the prime focus of 
Subaru Telescope, covering a $34\arcmin \times 27\arcmin$ field-of-view with a pixel scale of $0\arcsec.202$. 
The CCDs of Suprime-Cam were replaced to new fully depleted CCDs made by Hamamatsu Photonics 
in 2008 \citep{2008SPIE.7021E..52K} (see the next section). 
Because the practical ranges of redshifts observable for LAEs are limited by OH airglow lines, 
some dedicated NB filters, corresponding to $z=3.1, 4.8, 5.7$ and $6.6$ were manufactured 
\citep[e.g.,][]{2006ApJ...648....7K, 2011ApJ...734..119K, 2008ApJS..176..301O, 2010ApJ...723..869O, 2006PASJ...58..313S, 2005PASJ...57..165T}. 
Surveys performed using these NB filters revealed that 
 the Ly$\alpha$ LF exhibits almost no evolution from $z=3.1$ to $5.7$, 
whereas the Ly$\alpha$ LF at $z=6.6$ shows an apparent deficit, at least at its bright end, 
compared to that observed at $z=5.7$, while the rest-UV LFs at $z=6.6$ and $5.7$ are comparable. 
This result strongly supports the suggestion that the decline in the LAE 
luminosity function at $z>6$ was caused by the attenuation of the IGM rather than by the galaxy evolution.
The derived $L^*$ at $z=6.6$ is $30\%-60\%$ lower than that at $z=5.7$, indicating the luminosity attenuation 
toward higher redshift. On the other hand, \citet{2010ApJ...725..394H} reported that $L^*$ 
is unchanged from $z=5.7$ to $6.5$ based on their spectroscopically sample. Then, when and how the end of cosmic reionization 
was completed is still under debate.
Similarly, discovery of an LAE at $z\sim7.0$, IOK-1 at $z=6.96$, was made using the NB973 filter 
($\lambda_c =9755$\AA, $\Delta \lambda=200$\AA) imaging \citep{2006Natur.443..186I, 2008ApJ...677...12O}, 
revealing that the Ly$\alpha$ luminosity density gradually decreases from $z=5.7$ even to $7.0$.  
In fact, the number density of $z=7$ LAEs is only $18-36\%$ of the density at $z=6.6$.  
These results indicate a substantial transition in the IGM ionizing state 
during these early epochs. We might be observing the end of the cosmic reionization process.

To trace the neutral fraction as a function of redshift, it is necessary to survey for LAEs at higher-z. 
LAEs beyond $z=7$ have been surveyed by NIR instruments covering wide field-of-view and equipped with 
NB filters including the 4 m Mayall Telescope/NEWFIRM 
\citep{2010A&A...515A..97H, 2010ApJ...721.1853T, 2012ApJ...745..122K}. 
They reported that the Ly$\alpha$ LF has not evolved beyond $z=6$. 
These photometric LAE candidates have not yet been spectroscopically confirmed and possible presence of remaining contaminants
cannot be ruled out.  Therefore, photometric constraining of LFs at $z>7$ is not yet robust, 
and spectroscopic confirmation of the Ly$\alpha$ line is required.

Another effective method is to measure the fraction of Ly$\alpha$ emitting galaxies among LBGs in order 
to understand the reionization process. Sizable LBG samples at $z\sim 7$ found by 
the HST/WFC3 have been confirmed spectroscopically with the Keck/DEIMOS and the VLT/FORS2 
\citep{2010ApJ...725L.205F, 2012ApJ...744...83O, 2011ApJ...743..132P, 2012ApJ...744..179S, 2011ApJ...730L..35V}.
Their results suggest that the Ly$\alpha$ fraction at $z\sim 7$ could decline from $z \le 6$ 
\citep{2010MNRAS.408.1628S,2011ApJ...728L...2S} and it might be caused by the increased neutral fraction at $z\sim 7$. 
Hence, the current suggested redshift range of the end of the cosmic 
reionization from $\sim 6$ to $\sim 11$ indicates that the details of the cosmic reionization process remain obscure.


In this paper, we present the results of photometric and spectroscopic surveys for $z=7.3$ LAEs 
and use this sample to constrain the history of cosmic reionization.
The remainder of this paper is organized as follows. 
First we describe the imaging survey for $z=7.3$ LAEs in two blank fields in \S \ref{sec_imaging_observation}. 
We also explain the selection criteria of $z=7.3$ LAE candidates 
based on narrow-band and broad-band photometry in \S \ref{sec_photometric_candidates}.
Next, we present deep spectroscopic observations of our photometrically-selected LAE candidates in \S 
\ref{sec_spectroscopy}. 
We compare the estimated Ly$\alpha$ luminosity and SFR of 
$z=7.3$ LAEs with those of $z=6.6$ and $5.7$ LAEs 
and discuss constraints on cosmic reionization in \S \ref{sec_discussion}. 
In the last section, we summarize our results.

Throughout the present paper, we adopt concordance cosmology 
with $(\Omega_m, \Omega_\Lambda, h)=(0.3, 0.7, 0.7)$, 
\citep{2007ApJS..170..377S} and AB magnitudes with $2\arcsec$ diameter apertures 
for the NB1006 surveys unless otherwise specified.

\section{IMAGING OBSERVATIONS}\label{sec_imaging_observation}

\begin{figure}[t]
\epsscale{1.0}
\plotone{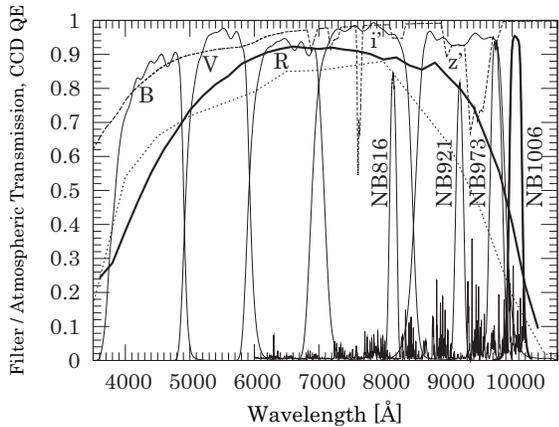}
\caption{Filter transmission curves of the Suprime-Cam  
  broad-band and narrow-band filters and the new NB1006 filter are shown with corresponding labels. 
  The spiky profile at the bottom represents the OH airglow lines. The OH airglow lines are not strong, 
  though not completely absent, at wavelength in the NB1006 filter.  
 The long-dashed curve at the top shows the atmospheric transmission. The short-dashed and thick solid 
 curves show the quantum efficiency of the previous MIT/LL CCDs and the new fully depleted CCDs, respectively. }
\label{fig_filter_system}
\end{figure}

The sensitivity at $1 \mu m$  of the previous Suprime-Cam was only about $20\%$. 
The CCDs of Suprime-Cam were upgraded to new red-sensitive ones in 2008 \citep{2008SPIE.7021E..52K}. 
As shown in Figure \ref{fig_filter_system}, these new CCDs have a quantum efficiency of $40\%$ at $1 \mu m$, 
twice that of the previous CCDs and enable an effective survey for LAEs beyond $z=7$. 
Actually, the survey using the new CCDs for LAEs at $z=7$ have obtained a deeper limiting magnitude \citep{2010ApJ...722..803O}
than that attained by the previous CCDs with same integration time.

We built a new NB filter, NB1006, to carry out a LAE survey at $z=7.3$ with the new red-sensitive Suprime-Cam. 
The transmission band of NB1006 filter is centered at $\lambda_c=10052$\AA, and has a FWHM $\Delta \lambda=214$\AA\, 
to cover $9943-10142$\AA, a modest, but not perfectly clean, OH band gap.
LAEs in the redshift range $z=7.261\pm 0.082$ can be detected with this filter (see Figure \ref{fig_filter_system}). 

We observed two sky regions, the Subaru Deep Field (SDF) and the Subaru XMM/Newton Deep survey Field
(SXDF) using the new red-sensitive Suprime-Cam and the new filter, NB1006.
The SDF is  centered at $13^h24^m21.^s4$, $+27^d29\arcmin23\arcsec(J2000)$ \citep{2004PASJ...56.1011K}.
The field has the deepest imaging data from those having a single pointing of Suprime-Cam. 
Deep broadband $B,V,R,i^{\prime},z^{\prime}$, and narrow-band NB816 ($\lambda_c = 8160$\AA, 
$\Delta \lambda_{FWHM}=120$\AA) and NB921 ($\lambda_c = 9196$\AA, 
$\Delta \lambda_{FWHM}=132$\AA) filter images were taken 
\citep{2004PASJ...56.1011K} and are made available for public access. 
The limiting magnitudes in a $2\arcsec$ aperture at $3\sigma$ are 
$B=28.45$, $V=27.74$, $R=27.80$, $i^{\prime}=27.43$, $z^{\prime}=26.62$, $NB816=26.63$, $NB921=26.54$, respectively. 
All of the images were convolved to a common seeing size of $0\arcsec.98$. 
These broad-band and narrow-band images can be downloaded from the official website of the SDF project
\footnote[10]{http://soaps.nao.ac.jp/}. 

The SXDF is centered at $02^h18^m00.^s0$, $-05^d00\arcmin00\arcsec(J2000)$ \citep{2008ApJS..176....1F, 2008ApJS..176..301O, 2010ApJ...723..869O}, and 
covers $\sim1.3$ square degree region of sky by five pointings of Suprime-Cam (SXDF-Center, North, South, East, West).
We chose a field to cover the southern half of SXDF-Center and the northern half of SXDF-South, 
where Spitzer SpUDS\footnote[11]{http://irsa.ipac.caltech.edu/data/SPITZER/docs/spitzermission/\\observingprograms/legacy/}, 
SEDS\footnote[12]{http://www.cfa.harvard.edu/SEDS/}, and HST CANDELS \citep{2011ApJS..197...35G, 2011ApJS..197...36K} 
data are also available. (see Figure.2 in \cite{2010ApJ...722..803O}).  
 The limiting magnitudes in the $2\arcsec$ aperture at $3\sigma$ are 
$B=28.09$, $V=27.78$, $R=27.57$, $i^{\prime}=27.62$, $z^{\prime}=26.57$, $NB816=26.55$, $NB921=26.40$ at the SXDF-Center and 
$B=28.33$, $V=27.75$, $R=27.67$, $i^{\prime}=27.47$, $z^{\prime}=26.39$, $NB816=26.65$, $NB921=26.40$ for SXDF-South. 
For SXDF, we matched the PSF of NB1006 images to the maximum PSF size of $0\arcsec.82$ as of the SXDF-Center and South.
The limiting magnitudes of available near infrared bands ($J$, $H$, and $K$) for SDF and SXDF are much shallower 
than those for optical bands and we did not use data from $J, H, K$ images for the photometric selection of candidates. 
The transmission curves of the optical filters used in the observation of SDF and SXDF are shown in 
Figure \ref{fig_filter_system}.
Spectroscopic standard stars GD153 and GD71 for SDF and SXDF were imaged during the 
observation to calibrate the photometric zeropoint.

The SDF and SXDF were observed using the NB1006 filter in February - April 2009 and October 2009 in $22.3$ and $14.8$ hour integrations, respectively. 
The observational log is summarized in Table \ref{table_observation_log}. 
We removed some images with poor quality from final analysis and the actual integration times used in the present paper 
are $17.2$ and $14.3$ hours for the SDF and SXDF, respectively. 
After masking out those regions that have low S/N ratios, the final effective areas of the SDF 
and SXDF images used were $855$ and $863$ arcmin$^2$, corresponding to 
the survey volumes of ($2.96$, $2.98$)$\times 10^5$ Mpc$^3$ for the SDF and SXDF, respectively.

\section{SELECTION OF PHOTOMETRIC CANDIDATES}\label{sec_photometric_candidates}

\subsection{Photometry}\label{subsec_photometry}

Source detection and photometry were carried out using SExtractor version 2.8.6 \citep{1996A&AS..117..393B}. 
We searched for objects having an area larger than $5$ contiguous pixels with a flux greater than $2\sigma$ excess 
over the sky surface brightness. Objects were chosen from the NB1006 image and their 
photometry was performed on the images in other wavebands at the same positions. 
The double-imaging mode of SExtractor was used for this task. The $2\arcsec$-diameter aperture 
magnitudes of detected objects were measured using the {\sf MAG\_APER} parameter and the total magnitudes 
were measured using {\sf MAG\_AUTO}. When the LAE candidates were selected, the aperture magnitude 
was used as the magnitude of object. 
Astrometric calibration was performed for images of NB1006 and the astrometric errors are $\sim 0\arcsec.2$. 

Bright stellar halos, saturated CCD blooming, and pixels of abnormally high or low 
flux count spikes were masked out in the SDF and SXDF images in all wavebands. 
The low $S/N$ regions at the edge of the image were also masked out. The final object catalog constructed from the 
NB1006 images contains $31,168$ objects for the SDF and $27,565$ objects for the SXDF 
down to NB1006$<24.83(5\sigma)$ for SDF and NB1006$<24.57(5\sigma)$ for SXDF.


\begin{figure}[t]
\epsscale{1.0}
\plotone{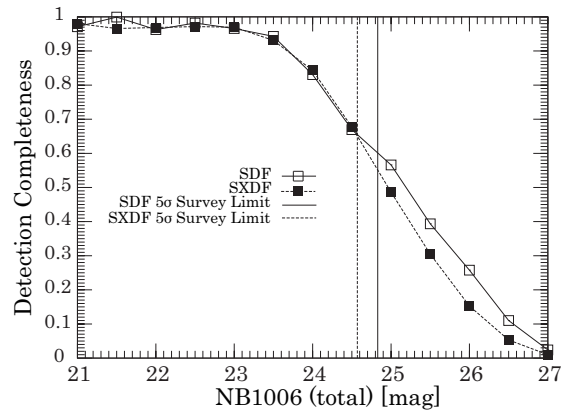}
\caption{The detection completeness of our SDF and SXDF NB1006 images per $0.5$ 
 mag bin. }
\label{fig_sdf_completeness}
\end{figure}

\begin{deluxetable*}{cccccc}
\setlength{\tabcolsep}{0.6cm} 
\tablecaption{Summary of the SDF and SXDF Observations}
\tablehead{  \colhead{Field} &  \colhead{$t_1$\tablenotemark{a}} & \colhead{the number of images} &  \colhead{$t_{exp}$\tablenotemark{b}} & \colhead{Date of Observations} &  \colhead{Seeing} \\ 
\colhead{}&  \colhead{[s]} &  \colhead{} &  \colhead{[s]} &  \colhead{}  &  \colhead{[arcsec]}} 

\startdata
    SDF & $500-1800$ & $19$ & $22094.5$ & 2009 February 25  & 0.8-1.1 \\ 
    SDF & $1200-1800$ & $15$ &  $22800$ & 2009 February 26  & 0.8-0.9 \\
    SDF & $730-1200$ & $7$ &  $7931.2$ & 2009 March 24  & 0.8-1.2\\
    SDF & $1200$ & $2$ &  $2400$ & 2009 April 1 & 1.4 \\
    SDF & $900-1200$ & $22$ &  $25200.8$ & 2009 April 2 & 0.8-1.2\\
    SXDF & $1200$ & $23$ &  $27600$ & 2009 October 15 & 0.4 \\
    SXDF & $1200$ & $21$ &  $25200$ & 2009 October 16  & 0.4
\enddata
\tablenotetext{a}{The integration time of one image.}
\tablenotetext{b}{The total integration time of the SDF is $80426.5$ s and the total integration time of the SXDF is $52800$ s.}
\label{table_observation_log}
\end{deluxetable*}

\subsection{Detection Completeness}\label{subsec_detection_completeness}

Photometric detection completeness in photometry is defined as the ratio of the number of 
objects detected by the photometry to that of the objects actually present in the universe. 
In order to evaluate the detection completeness of our detections down to the limiting magnitude of our NB1006 images, 
the {\sf starlist} and {\sf mkobjects} tasks in IRAF were used.   We created a sample $20,000$ artificial stellar objects having 
the same PSF as on the NB1006 images with a random spatial distribution over blank regions of the images and 
with magnitudes ranging from $21$ to $27$ mag.
The SExtractor was run for source detection on these mockup image data in exactly the same way as we did in our actual photometry. 
We used the {\sf ASSOC}iation mode of the SExtractor to avoid sky noise not located on the artificial 
stellar objects. We repeated this procedure five times and averaged the obtained detection completeness. 
The results are shown in Figure \ref{fig_sdf_completeness}. It is likely that the detection completeness exceeds 
$90 \%$ at NB1006 $<23.7$, and more than $50 \%$ even at the 5 sigma limiting magnitudes of 
$NB1006=24.83, 24.57$, for SDF and SXDF, respectively. 
This detection completeness factors were used to correct the number and luminosity densities 
of $z=7.3$ LAEs discussed in \S \ref{sec_discussion}.

\subsection{Selection of $z=7.3$ LAE candidates}\label{subsec_selection_candidate}

We select LAE candidates at $z=7.3$ as those objects showing significant flux in NB1006 but not detected in any of  
the bluer bands. 
The ($z^{\prime}-NB1006, NB1006$) color magnitude distribution of all the detected objects in the SDF and SXDF are shown 
in Figures \ref{fig_cmd_sdf} and \ref{fig_cmd_sxdf}, respectively. 
The average $z^{\prime}-NB1006$ color excess in the magnitude range $22 < NB1006 < 24$ was about $+0.2$. 
The dashed curves in these figures show $3\sigma$ boundaries of $z^{\prime}-NB1006$ color distribution defined by the equation 
$\pm 3 \sigma_{BB-NB}=-2.5 \log_{10} (1\mp (f^2_{3\sigma NB} + f^2_{3\sigma BB})^{0.5}/f_{NB})$.

As the diagram indicates, if $z^\prime - NB1006 \ge2.6$ is used as the color selection criterion, then $z=7.3$ LAEs can be identified. 
It should be noted that the NB1006 band pass is located at the red edge of the CCD sensitive 
wavelengths; therefore, the UV continuum redder than Ly$\alpha$ emission is hardly detected 
even in the $z^{\prime}$-band.
Hence, we applied the following selection criteria to the detected objects in the same way as \citet{2008ApJ...677...12O}:


\begin{figure}[t]
\epsscale{1.0}8
\plotone{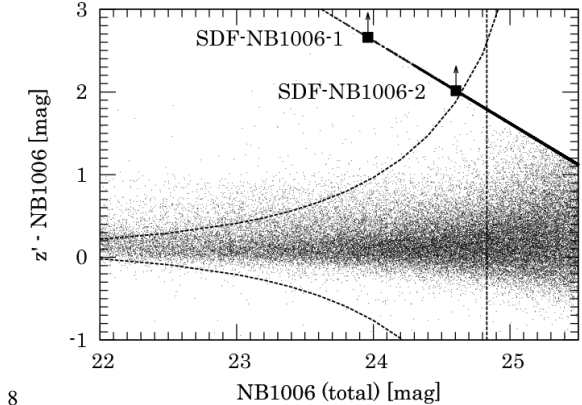}
\caption{$z^{\prime}-NB1006$ color as a function of NB1006 magnitude of the objects detected in the 
 SDF (shown by dots). The vertical line is the limiting magnitude of the SDF, $24.83$ mag. 
 The diagonal dashed line is the $3\sigma$ limit of the $z^{\prime}$ magnitudes in $2\arcsec$ aperture. 
 The points on this diagonal dashed line are the colors which are converted to the 
 $3\sigma$ limits of $z^{\prime}$ magnitudes in $2\arcsec$ aperture.
 The curve shows the $3\sigma$ error track of $z^{\prime}-NB1006$ color. 
 The squares are the $z=7.3$ LAE candidates, SDF-NB1006-1 and SDF-NB1006-2.}
\label{fig_cmd_sdf}
\end{figure}

\begin{enumerate}
  \setlength{\parskip}{0cm}
  \setlength{\itemsep}{0cm}
 \item $NB1006>5\sigma : \\NB973, NB921, NB816, i^{\prime} ,R ,V ,B < 3\sigma, z^{\prime}-NB1006\ge 2.6$
 \item $NB1006>5\sigma : \\NB973, NB921, z^{\prime} ,NB816 ,i^{\prime} ,R ,V ,B < 3\sigma$
\end{enumerate}

\noindent where all magnitudes were measured as $2\arcsec$-diameter aperture magnitudes, {\sf MAG\_APER}.

Color criterion, $z^{\prime} - NB1006 > 2.6$, was determined by the model spectrum of a galaxy 
at $z=7.3$. We created a spectral energy distribution (SED) of a starburst galaxy using stellar population synthesis model, 
GALAXEV \citep{2003MNRAS.344.1000B}, with a metallicity of $Z=Z_{\odot}=0.02$, an age of $t=1$ Gyr, a Salpeter 
initial mass function with lower and upper mass cutoffs of $m_L=0.1 M_{\odot}$ and $m_U=100 M_{\odot}$, and 
an exponentially decaying star formation history for $\tau =1$ Gyr. These parameters were selected to be the same 
as those used in \citet{2005PASJ...57..165T, 2008ApJ...677...12O}, for consistency. 
The absorption by IGM was applied to 
the generated spectrum using the model of \citet{1995ApJ...441...18M}. 
The flux of a Ly$\alpha$ emission line with EW$_0 =0$\AA\ was added to the SED at $(1 + 7.26) 1216$\AA. 
In other words, we create a SED of LBG with no Ly$\alpha$ emission line. 
Hence, the presently adopted color criterion, $z^\prime - NB1006 > 2.6$, corresponds to the
rest-frame equivalent width of Ly$\alpha$ of EW$_0 > 0$.
Moreover, the photometric error may cause interlopers at lower-$z$ to happen to satisfy our 
color selection criteria. We carried out the following test in the same way as did in 
\cite{2011ApJ...737...90B, 2011MNRAS.411...23W}. First, we choose brighter sources in magnitude 
range of $22.5 > NB1006 > 23.5$ from our two fields and then dim these bright sources to match 
to the same magnitudes as of our LAE candidates by scaling the optical BB and NB fluxes. 
We then distribute these objects over the real images. 
The number of these artificial objects is the same as the number of observed objects 
in the magnitude range from $NB1006 < 23.5$ down to each NB1006 limiting magnitude. 
We applied exact the same detection and selection criteria as the candidate selection. 
As a result, we found no objects happen to come into our criteria, suggesting that 
our selection criteria can robustly avoid any nearby red and line emitting interlopers.

\begin{figure}[t]
\epsscale{1.0}
\plotone{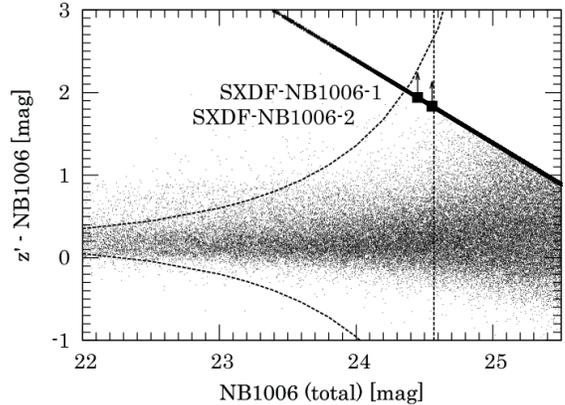}
\caption{Same as Figure \ref{fig_cmd_sdf}, but the filled squares and dots are the candidates and objects detected in the SXDF, respectively. 
    The objects in two fields in SXDF, 
    SXDF-C and SXDF-S which have different $z^{\prime}$ limiting magnitudes are plotted in this same diagram. 
    The fainter magnitudes of objects than $z^{\prime}$ limiting magnitude of each fields are converted to the shallower 3$\sigma$ 
    limits (SXDF-S).}
\label{fig_cmd_sxdf}
\end{figure}

\begin{figure*}
\epsscale{1.0}
\plotone{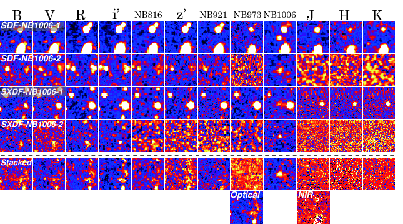}
\caption{Multi-waveband images of the $z=7.3$ LAE candidates in the SDF and the SXDF. The postage 
 stamps have an area of $10\arcsec \times10\arcsec$. The optical images are the archival data, 
 which were observed in  the SDF and the SXDF projects. 
 The near-infrared images were observed by the UKIRT/WFCAM \citep{2007ApJ...660...72H}.
 Second row from the bottom: Centered-staked postage-stamp images for LAE candidates. 
 Stacks have no flux in the optical- and NIR- filters. 
 Bottom row: Centered-stacked postage-stamp images using only optical- and NIR- filters.}
\label{fig_candidate}
\end{figure*}

As a consequence of selection, we found no candidates meeting only criterion (1) and two objects satisfying both 
criteria (1) and (2) in each field.
The images and photometric properties of these LAE candidates are listed in Figure 
\ref{fig_candidate}, and Table \ref{table_candidate}. 
We detected no flux even in the stacked optical and NIR images of all
the candidates, as shown in the bottoms of Figure \ref{fig_candidate}.


\begin{deluxetable*}{cccccc}
\setlength{\tabcolsep}{0.4cm} 
\tablecaption{Photometric Properties of $z=7.3$ Ly$\alpha$ Emitter Candidates}
\tablehead{\colhead{Name} & \colhead{RA} & \colhead{DEC} & \colhead{$z^{\prime}$\tablenotemark{a}} & \colhead{NB1006 $_{APER}$} & \colhead{NB1006 $_{AUTO}$}\\ 
\colhead{} & \colhead{(J2000)} & \colhead{(J2000)} & \colhead{[mag]} & \colhead{[mag]} & \colhead{[mag]}} 

\startdata
SDF-NB1006-1 & $13:24:39.993$ & $+27:39:40.67$ & $>26.62$ & $23.96\pm0.03$ & $23.62\pm0.04$ \\
SDF-NB1006-2 & $13:24:35.418$ & $+27:27:27.81$ & $>26.62$ & $24.61\pm0.05$ & $24.21\pm0.05$\tablenotemark{b}  \\
SXDF-NB1006-1 & $02:18:50.657$ & $-05:24:40.25$ & $>26.39$ & $24.45\pm0.05$ & $23.81\pm0.05$ \\
SXDF-NB1006-2 & $02:18:56.536$ & $-05:19:58.92$ & $>26.39$ & $24.56\pm0.06$ & $24.26\pm0.07$
\enddata
\tablenotetext{a}{The $3\sigma$ limiting magnitudes of the candidates in $z^{\prime}$-band were derived from the $2\arcsec \phi$-aperture photometry.}
\tablenotetext{b}{The NB1006 magnitude of the SDF-NB1006-2 was derived by $3\arcsec$ diameter aperture photometry instead of 
{\sf MAG\_AUTO} to avoid contamination from the flux of a nearby object.}
\label{table_candidate}
\end{deluxetable*}

\section{SPECTROSCOPY}\label{sec_spectroscopy}

We carried out optical deep spectroscopy for three of our four photometric LAE candidates. 
 We used the Faint Object Camera and Spectrograph (FOCAS) in Multi-Object Spectroscopy (MOS) mode mounted on the 
Subaru telescope on 2010 March 18, 19, and 21 to follow the two candidates in the SDF, SDF-NB1006-1 
and SDF-NB1006-2. 
The observations were made with a $555$ mm$^{-1}$ grating and an O58 order-cut filter, giving spectral 
coverage of $7500-10450$\AA\, with a dispersion of $0.74$\AA pix$^{-1}$. The $0\arcsec.8$-wide slit used gave a 
spectroscopic resolution of $5.7$\AA\, (R$\sim 1500$) which is sufficient to distinguish $[$O {\sc ii}$]$ doublet lines 
from possible contaminant galaxies at z=$1.7$. 
However, it is likely that $[$O {\sc ii}$]$ doublet lines are not resolved in this resolution 
if the emission lines are too faint or are affected by OH sky lines. 
The observing nights were photometric, with good seeing of $\sim 0\arcsec.6 - 1\arcsec.0$. 
Because no emission line feature was visible for the brighter candidate, SDF-NB1006-1,  
in the entire transmission range of the NB1006 filter in 5 hours integration, this object was abandoned.
We switched on to the second candidate,  SDF-NB1006-2, and secured spectra for a total integration time of 11.5 hours.

In addition to the observations made in March, we observed SDF-NB1006-2 for a half night on 2010 May 5. 
The seeing $0\arcsec.8-1\arcsec.0$ was somewhat poor but the weather condition was photometric.
The setup of the instrument was exactly the same as that in March. 
The integration time in May was $4.5$ hours, and total integration time for the SDF-NB1006-2 spectroscopy 
is $16$ hours.


\begin{deluxetable*}{cccccc}
\setlength{\tabcolsep}{0.5cm} 
\tablecaption{the Summary of the Follow Up Spectroscopy}
\tablehead{\colhead{Target}  & \colhead{$t_1$\tablenotemark{a}} & \colhead{the number of image} & \colhead{$t_{exp}$}  & \colhead{Date of Observations} & \colhead{Instrument}  \\ 
\colhead{} & \colhead{[s]} & \colhead{} & \colhead{[s]} & \colhead{} & \colhead{}} 

\startdata
    SDF-NB1006-1 & $1800$ & $10$ & $18000$ & 2010 March 18-21  & Subaru/FOCAS \\ 
    SDF-NB1006-2 & $1800$ & $23$ &  $41400$ & 2010 March 18-21  & Subaru/FOCAS \\
    SDF-NB1006-2 & $1800$ & $9$ &  $16200$ & 2010 May 5  & Subaru/FOCAS \\
    SXDF-NB1006-2 & $1800$ & $7$ &  $12600$ & 2010 October 10  &  Keck/DEIMOS
\enddata
\tablenotetext{a}{The integration time of one image.}
\label{table_spectroscopy_log}
\end{deluxetable*}

\begin{figure}[t]
\begin{flushleft}
\epsscale{1.0}
\plotone{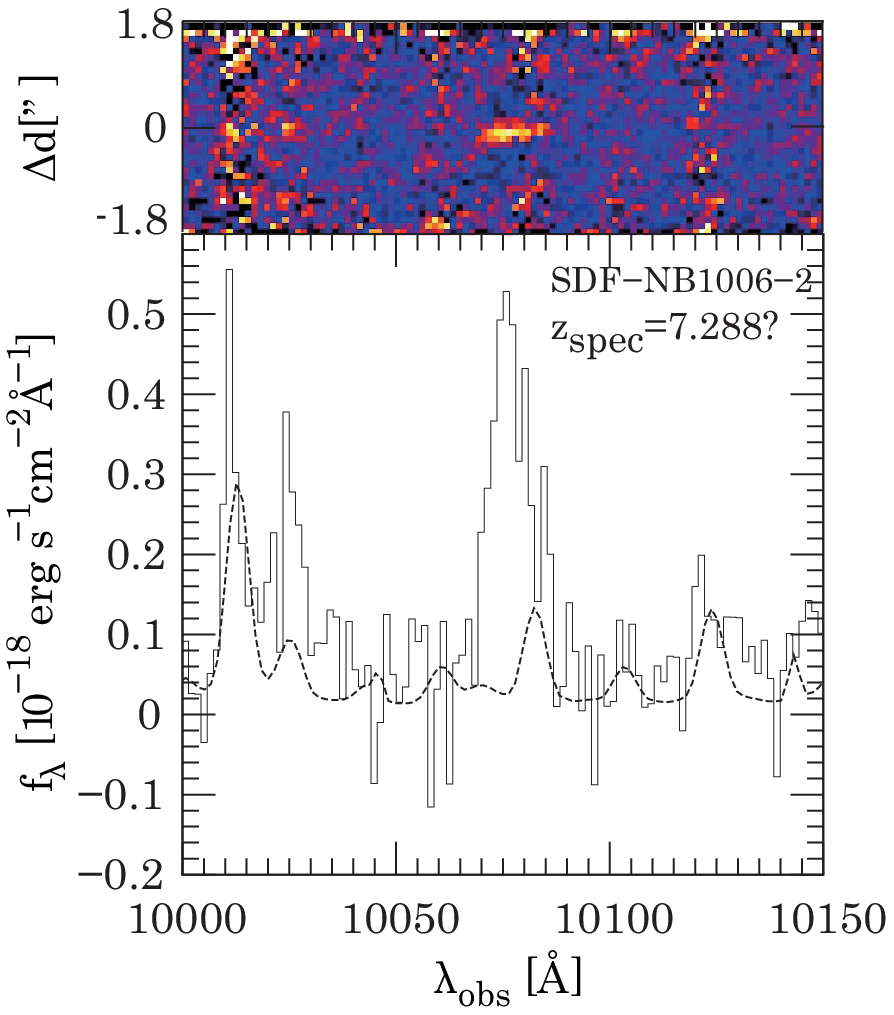}
\caption{Top panel: Composite two-dimensional spectrum of 
  SDF-NB1006-2. Bottom panel: Combined spectrum of SDF-NB1006-2. 
  The dashed line represents the OH sky line. A prominent emission line is seen at $10075.8$\AA\, corresponding to $z=7.288$ 
  if this line is Ly$\alpha$. The weighted skewness of the line is $4.59\pm1.25$. 
  This value indicates that this emission line is asymmetric because it is greater than the threshold ($S_{\omega}=3$) 
  used in the LAE studies at $z=5.7$ and $6.6$ \citep{2006ApJ...648....7K}. The $S/N$ ratio of the line is $\sim 8.3$. 
  The dashed line represents OH sky line.}
\label{fig_spec_sdf2}
\end{flushleft}
\end{figure}

Each of the $30$ minute exposures was taken by dithering the telescope pointing 
along the slit by $\pm 1\arcsec.0$ under good seeing conditions and $\pm 1\arcsec.5$ under moderate seeing 
conditions to enable more accurate sky subtraction. The standard star Feige 34 and Hz 44 were taken at 
the beginning and end of each observed night except on May 5 when only Feige 34 was observed \citep{1990ApJ...358..344M}. 
Feige 34 and Hz 44 were used for flux calibration of May and March data, respectively.
The object and standard star data were reduced in a standard manner with IRAF.
Also, we evaluated the slit loss for the Ly$\alpha$ flux measurements calculated from the spectra. 
The targets were assumed to be almost perfectly centered in the slit and point sources. 
The slit loss was calculated by putting a Gaussian profile with FWHM derived from the average seeing size 
in the FOCAS spectroscopic observation on the $0\arcsec.8$-wide slit. 
The slit loss of the FOCAS spectroscopy was evaluated to be $\sim 23$\%. 
The resultant slit loss was corrected for the spectroscopically measured Ly$\alpha$ for the object. 
The stacked one- and two- dimensional spectra of SDF-NB1006-2 are shown in Figure  
\ref{fig_spec_sdf2}. A prominent emission line is seen at $10075.8$\AA. 

We also used DEep ImagingMulti-Object Spectrograph (DEIMOS) in MOS mode at the 
Nasmyth focus of the Keck II telescope to take spectroscopy for the candidate, SXDF-NB1006-2
as described in (Ota et al. in prep). 
The seeing was $\sim 0\arcsec.6-1\arcsec.0$. We used the GG495 filter and the
830 lines mm$^{-1}$ grating, which was tilted to place a central wavelength of $9000$\AA\, on the
detectors. This configuration provided spectral coverage between $7000$\AA\, and $10400$\AA. 
The spatial pixel scale was $0\arcsec.1185$ pix$^{-1}$, and the spectral dispersion was 
$0.47$\AA pix$^{-1}$. The slit widths were $1\arcsec.0$, which gives the spectral resolution of $3.3$\AA. 
The total integration time of DEIMOS is $3.5$ hours. 

The reduction of the spectra for SXDF-NB1006-2 was performed using spec2d IDL
pipeline developed by the DEEP2 Redshift Survey Team \citep{2003SPIE.4834..161D}. 
The final one-dimensional data output from the spec2d, however, show apparent over-subtraction of the foreground OH 
sky lines. In order to solve this difficulty, we modified the code in the process of calibrating 
the wavelength using arc lamp and subtracted OH sky lines carefully in a manner similar to that 
employed in the FOCAS data reduction with IRAF. 


\begin{flushright}
\begin{figure}[t]
\epsscale{1.0}
\plotone{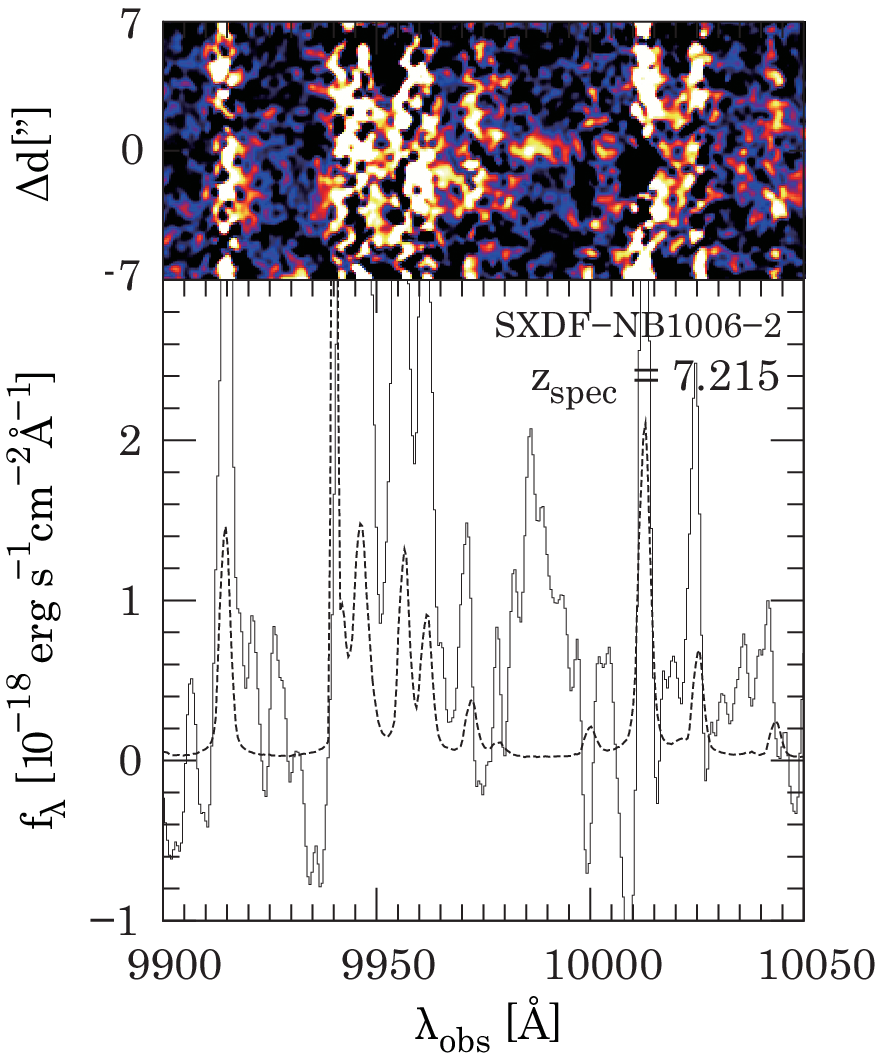}
\caption{ Top panel: Combined spectrum of the SXDF-NB1006-2. 
  Bottom panel: Combined spectrum of the SXDF-NB1006-2.
  A line is visually identified at $9987.6$\AA. The $S/N$ ratio of the line is $\sim 4.5$. Both spectra are smoothed by being 
  convolved with the $\sigma = 1.5$ [pixel] Gaussian function. The weighted skewness line is $4.90\pm0.86$. 
  This value indicates that this emission line is asymmetry because it is greater than the threshold ($S_{\omega}=3$) 
  used in the LAE studies at $z=5.7$ and $6.6$ \citep{2006ApJ...648....7K}. The dashed line represents OH sky line.}
\label{fig_spec_sxdf2}
\end{figure}
\end{flushright}

Although we also observed the standard star Feige 110 for flux calibration with DEIMOS, 
these data were not reduced due to a computational problem. 
Hence, the flux of the SXDF-NB1006-2 was calibrated using the spectra of five late-type stars observed simultaneously 
in the same slit mask mounted on DEIMOS. 
We compared their spectroscopically measured fluxes with those expected from 
their photometry convolved with the filter (NB1006) transmission and derived the zero-point for the spectroscopic flux calibration. 
The statistical fluctuation in the above flux calibration with five late-type stars was used as the estimated flux error of the SXDF-NB1006-2. 
The slit loss has already been corrected in this calibration. 
The stacked one- and two- dimensional spectrum of the SXDF-NB1006-2 are shown in Figure \ref{fig_spec_sxdf2}.

We measured the weighted skewness of the detected emission lines, which can be used as a criterion to identify 
high-$z$ Ly$\alpha$ emission lines, \citep{2006ApJ...648....7K}. The weighted skewness was calculated 
in the wavelength range where the flux is higher than $10$\% of the peak value, ranging 
from $10068$\AA\, to $10088$\AA, and from $9980$\AA\, to $9998$\AA, for SDF-NB1006-2 and SXDF-NB1006-2, respectively.
They were $S_\omega =4.59\pm1.25$ for SDF-NB1006-2 and $4.90\pm0.86$ for SXDF-NB1006-2. 
These measured skewness values satisfy the usually adopted criterion for the presence of asymmetry of high-$z$  
Ly$\alpha$ emission, $S_\omega> 3$ for typical LAEs at $z>5.7$. 
Note however that the red side of emission line in the SDF-NB1006-2 spectrum has an overlap with 
a weak OH sky line, and the derived skewness value might be overestimated.

We had no chance to observe the remaining fourth candidate in the SXDF, SXDF-NB1006-1 during our allocated time.  
The spectroscopic properties of the line emitting objects are summarized in Table \ref{table_spectroscopy}.

\begin{deluxetable*}{cccccccccc}
\setlength{\tabcolsep}{0.01cm} 
\tabletypesize{\scriptsize}
\tablecaption{Spectroscopic Properties of Line Emitting Objects}
\tablehead{\colhead{Name} & \colhead{Redshift} & \colhead{$f^{Ly\alpha}$\tablenotemark{a}} & \colhead{$L^{Ly\alpha}$} & \colhead{SFR$^{Ly\alpha}$\tablenotemark{b}} & \colhead{$M_{UV}$\tablenotemark{c}} & \colhead{SRF$^{UV}$\tablenotemark{d}} & \colhead{FWHM\tablenotemark{e}} & \colhead{$S_{\omega}$} & \colhead{EW$_0$\tablenotemark{f}}  \\ 
\colhead{} & \colhead{} & \colhead{[10$^{-17}$erg s$^{-1}$cm$^{-2}$]} & \colhead{[10$^{43}$erg s$^{-1}$]} & \colhead{[M$_{\odot}$yr$^{-1}$]} & \colhead{[mag]} & \colhead{[M$_{\odot}$yr$^{-1}$]} & \colhead{[\AA]} & \colhead{[\AA]} & \colhead{[\AA]}}

\startdata
SDF-NB1006-2 &  $7.288?$\tablenotemark{g} & $0.91\pm0.13$ &  $0.57\pm0.15$ &  $5.2\pm1.1$ & $-23.79\pm0.04$ & $1646\pm54?$\tablenotemark{h} &  $13.4$ &  $4.59\pm1.25$ & $1.99\pm0.37$ \\
SXDF-NB1006-2 & $7.215$ & $1.9^{+2.5}_{-0.9}$ &  $1.2^{+1.5}_{-0.6}$ &  $11^{+15}_{-5.3}$ & $-22.4^{+\infty}_{-0.4}$ & $437^{+190}_{-437}$ &  $11.5$ & $4.90\pm0.86$\tablenotemark{i} & $15.4^{+\infty}_{-4.7}$
\enddata
\tablenotetext{a}{The observed line flux corresponds to the total amount of the flux within the line profile. A correction was made for slit loss (see \S \ref{sec_spectroscopy}). }
\tablenotetext{b}{The SFRs were derived from $L^{Ly\alpha}$: $ SFR^{Ly\alpha} = 9.1\times10^{-43} L^{Ly\alpha} M_{\odot} yr^{-1}$ \citep{1998ARA&A..36..189K}}
\tablenotetext{c}{The absolute magnitude of the UV continuum were derived from the UV fluxes  $f^{UV} = f_{NB1006} - f^{Ly\alpha}$ using equation (1) of \citet{2011ApJ...734..119K}.}
\tablenotetext{d}{The SFRs were derived from $M_{UV}$: $SFR^{UV} = 1.4 \times 10^{-28} L_\nu (UV) M_\odot yr^{-1}$ \citep{1998ARA&A..36..189K}}
\tablenotetext{e}{The FWHMs (in observed frame) were derived from the profile fitted by the Gaussian function, since the actual spectra are noisy.}
\tablenotetext{f}{The rest-frame equivalent widths of Ly$\alpha$ emission were derived from the combination of $f^{Ly\alpha}$ and 
$M_{UV}$.}
\tablenotetext{g}{The NB1006 magnitude of the SDF-NB1006-2 fluctuated approximately $\pm 0.8$ mag, and it might be a QSO/AGN at $z=1.703$ or $z=7.288$. If SDF-NB1006-2 is a [O{\sc ii}] emitter, the $L_{[OII]}$ and $SFR_{[OII]}$ are $1.77\times10^{41} 
[erg s^{-1}]$ and $2.48 [M_{\odot}yr^{-1}]$, respectively.}
\tablenotetext{h}{This high UV SFR of SDF-NB1006-2 may originate from the central QSO/AGN engine. }
\tablenotetext{i}{The weighted skewness of the SXDF-NB1006-2 were derived from the smoothed emission line.}
\label{table_spectroscopy}
\end{deluxetable*}

\section{DISCUSSION} \label{sec_discussion}

\subsection{Check for the Variability of the Candidates}\label{subsec_variability}

Since the NB1006 imaging observations we used for detecting our $z=7.3$ LAE candidates were made a few years 
after the imaging observations for other bands, some of the objects which are only visible in NB1006 
could be variable objects, such as galaxies with supernovae or enhanced active galactic nuclei (AGNs), 
which happened to increase their luminosities during our NB1006 imaging observations. 
Since our SDF imaging was made in three epochs, we created three stacked NB1006 SDF images, one for February 2009, 
one for March 2009, and one for April 2009 to check for any variability. 
The SXDF observations were taken on two successive nights on October 15 and October 16. 
Although the time interval is short for SXDF, we also examined possible photometric variability. 

\begin{figure}[t]
\epsscale{1.0}
\plotone{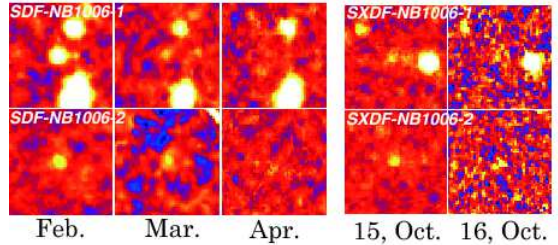}
\caption{The images of the $z=7.3$ LAEs candidates in the divided images for each 
 observational epoch. The PSF sizes are matched to $0\arcsec.98$ in SDF and $0\arcsec.82$ in SXDF.}
\label{fig_variability}
\end{figure}

Objects were chosen from the stacked NB1006 image in all epochs and 
photometry was performed on the images in the individual epochs at exactly 
the same positions as the objects in the NB1006 filter using the double-imaging mode 
of SExtractor. The results of independent brightness measurements at different epochs of observation are summarized 
in Table \ref{table_variability} and shown as postage stamp images of the individual epochs are shown in Figure \ref{fig_variability}.

As shown in the table, the magnitudes of candidates SDF-NB1006-1, and 2 vary by more than 
$\sim 0.5$ mag over two months, while the average magnitude inconsistency of other objects in the SDF NB1006 images 
which have the similar magnitudes as the LAE candidates is less than $0.1\pm0.6$ mag.  
In particular, SDF-NB1006-1, the brightest among our $z = 7.3$ LAE
candidates, appears to show a significant luminosity decline of $1.8$ mag
as shown in Figure \ref{fig_variability}. The brightness change corresponds to $7 \sigma_{phot}$,
where $\sigma_{phot}$ is the $1 \sigma$ photometric error of the object.
This amplitude of variability is roughly consistent with the J-band light-curve of a type Ia supernova, 
depending on the phase of its explosion \citep{2011AJ....141...19B}.
We conclude that SDF-NB1006-1 is a variable object which brightened suddenly in 
February and then gradually faded. It may be a galaxy with a supernova, which is consistent with the fact that 
no obvious emission line was confirmed in the FOCAS spectroscopy. 

The possible magnitude variation of SDF-NB1006-2, which shows an emission line 
with an asymmetric profile is a bit more enigmatic.  
The emission line indicates that it could be a LAE at $z=7.288$ considering its high weighted skewness ($S_\omega = 4.59 \pm 1.25$). 
On the other hand, because the NB1006 magnitude of this object changed by $\pm 0.8-0.9$ mag ($3.3-7.9 \sigma_{phot}$) 
over about one month, it could also be an AGN or a QSO. 
This suggestion might be supported by the high UV magnitude of SDF-NB1006-2 derived by 
subtracting the Ly$\alpha$ flux from the total flux in the NB1006 magnitude. 
Considering that the line width is narrower than that of typical QSOs, it is unlikely to be a QSO. 
The Ly$\alpha$ flux emitted from the broad line region (BLR) in AGNs/QSOs varies slightly, 
while that of $[$O {\sc ii}$]$ irradiated from the narrow line region (NLR) does not change \citep{2007ApJ...659..997K}. 
Besides, the change of UV continuum flux radiated from the BLR is investigated by the structure function of 
the photometric variability of AGNs/QSOs \citep{2005AJ....129..615D}, and is $\sim 0.55$ mag for a {\it year}. 
The structure function is the ensemble average of QSO variability as function of time and has been used as a tool to 
characterize the QSO variability. 
Since the UV continuum flux of the SDF-NB1006-1 is only $\sim70$\% of the NB1006 flux, its large variability might not be due 
to AGNs/QSOs, though the variation of the investigated structure function in the literature is relatively large. 
Hence, this object might be an AGN/QSO at $z=7.288$ if this line is Ly$\alpha$, or at $z=1.703$ if that is $[$O {\sc ii}$]$, 
rather than a LAE, nevertheless the large amplitude of the variability cannot be explained adequately. 

By contrast, the magnitude of the candidates in the SXDF, SXDF-NB1006-1 and 2, did not show 
any significant variation that would lead us to suspect their AGN nature. 
The brightness changes of SXDF-NB1006-1 and 2 correspond to $0.2$ mag ($1.1 \sigma_{phot}$) and $0.42$ mag 
($2.2 \sigma_{phot}$), respectively. 
The reality of the SXDF-NB1006-2 at $z=7.215$ is also supported by this result of checking variability. 

Moreover, we investigate the possibility that the detected lines could be H$\alpha$, H$\beta$, or $[$O {\sc iii}$]$  
in terms of metallicities estimated from diagnostic line ratios. 
The same case test for $z\sim 6.9-7.2$ LBGs was also discussed in \citet{2012ApJ...744...83O}. 
Any other emission line features were not observed in the spectra; therefore the metallicity is only given 
as an upper limit by using the flux limit. If the detected lines were H$\beta$, 
$[$O {\sc iii}$]$$\lambda 5007$ line would fall at $10378$\AA\, and $10288$\AA\, for SDF-NB1006-2 and SXDF-NB1006-2.  
The upper limits of $[$O {\sc iii}$]$$\lambda5007/$H$\beta$ line ratios are calculated to be $\le 0.1-0.4$, 
corresponding to the metallicities of $12+$log(O/H) $\ge 9.0-9.5$ based on the Figure 17 of \citet{2006A&A...459...85N}. 
In the case of such high metallicities, they should be very massive and bright taking account the mass-metallicity relation. 
However, they are not detected in our deep broadband images. If the detected lines were $[$O {\sc iii}$]$$\lambda 5007$, 
the H$\beta$ line would be detected at $9782$\AA\, and $9697$\AA\, for SDF-NB1006-2 and SXDF-NB1006-2, and  
the lower limits of $[$O {\sc iii}$]$$\lambda5007/$H$\beta$ are given as $\ge 11$ and $\ge 2$, respectively. 
The detected line of the SDF-NB1006-2 is unlikely to be $[$O {\sc iii}$]$, because  no outlier with 
$[$O {\sc iii}$]$$\lambda5007/$H$\beta > 10$ in \citet{2006A&A...459...85N}. 
The deep FOCAS spectroscopy for SDF-NB1006-2 completely rules out the possibility of other lines.
On the other hand, the lower limits for SXDF-NB1006-2 corresponds to subsolar oxygen abundances, which may be 
typical of low-mass galaxies. Hence, the non-detection of H$\beta$ cannot strongly rule out the possibility of the detected line 
in the SXDF-NB1006-2 spectrum being $[$O {\sc iii}$]$$\lambda 5007$. 
The possibility of H$\alpha$ can be ruled out because either $[$O {\sc iii}$]$ or H$\beta$ is expected to be 
simultaneously detected at the observed wavelength for both targets.
The possibility that the detected line is the unresolved $[$O {\sc ii}$]$ doublet can be ruled out for SXDF-NB1006-2 
due to the spectral resolution of DEIMOS. On the other hand, for SDF-NB1006-2, we cannot deny the possibility of $[$O {\sc ii}$]$.

After all, we could completely rule out the possibility of the detected emission lines 
being other nebular emission lines, i.e. H$\alpha$, H$\beta$, and $[$O {\sc iii}$]$. One possibility that the 
detected line of SXDF-NB1006-2 is H$\beta$ could not be completely ruled out based on the diagnostic line ratios. 
However, we should note that the asymmetric line profile strongly implies the line to be Ly$\alpha$ emission. 
The SDF-NB1006-1 and 2, which have significant variability amplitude more than
$3 \sigma_{phot}$, are concluded as variable objects.
We conclude that the SXDF-NB1006-2 is a plausible LAE at $z=7.215$, but the SDF-NB1006-2 
is likely to be a AGN/QSO either at $z=7.288$ or at $z=1.703$. 
Our results indicate that checking the time variability is crucial to strictly verify the LAE photometric candidates 
if they have been detected in only a single band. The checking the variability is also attempted in \citet{2011Natur.469..504B}. 
This check implies a possibility that two of the LAE candidates in the SDF 
could be transient objects such as an AGN/QSO or supernova, even though an asymmetric line was detected. 
However, we should note that the time span in SXDF was only one day delay which was too short to 
completely trace the typical variability.


\begin{deluxetable*}{cccc}
\tablecaption{The variability of the $z=7.3$ LAE Candidates}
\tablehead{\colhead{SDF} & \colhead{Feburuary} & \colhead{March (Mar.-Feb.)} & \colhead{April (Apr.-Feb.)} \\ 
\colhead{} & \colhead{[mag]} & \colhead{[mag]} & \colhead{[mag]} } 
\startdata
SDF-NB1006-1 &  23.67$\pm$0.027 &  24.31$\pm$0.097 (0.64$\pm$0.12, $5.3 \sigma_{phot}$) &  25.49$\pm$0.23 (1.82$\pm$0.26, $7.0 \sigma_{phot}$) \\
SDF-NB1006-2 &  24.53$\pm$0.051 &  23.75$\pm$0.059 (-0.87$\pm$0.11, $7.9 \sigma_{phot}$) &  25.35$\pm$0.20 (0.82$\pm$0.25, $3.3 \sigma_{phot}$) \\ \hline
SXDF & 15, October & 16, October (16, Oct. - 15, Oct.) &  \\
 & [mag] & [mag] &  \\ \hline
SXDF-NB1006-1 &  24.45$\pm$0.055 &  24.65$\pm$0.12 (0.20$\pm$0.18, $1.1 \sigma_{phot}$) \\
SXDF-NB1006-2 &  24.44$\pm$0.054 &  24.86$\pm$0.14 (0.42$\pm$0.19, $2.2 \sigma_{phot}$)
\enddata
\tablecomments{All the magnitudes are calculated by $2\arcsec$-aperture photometry. 
The aperture magnitudes ($2\arcsec$) at each epoch, the difference from  the magnitude at  the first observed epoch, and 
the variability significance compared with the photometric error $\sigma_{phot}$ are listed. 
The PSF sizes are matched to $0\arcsec.98$ in the SDF and $0\arcsec.82$ in the SXDF.}
\label{table_variability}
\end{deluxetable*}

 
 \begin{figure*}[t]
\epsscale{0.6}
\plotone{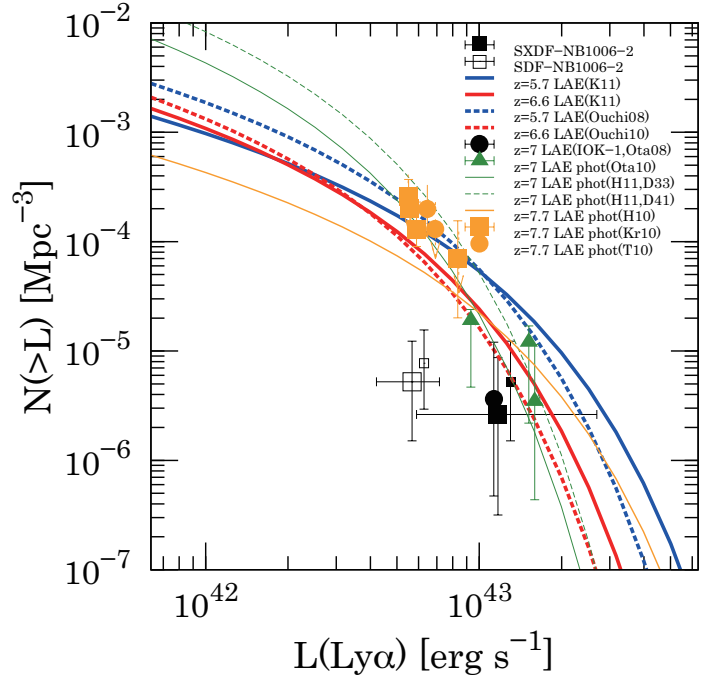}
\caption{Cumulative Ly$\alpha$ LFs at $z\le7$ and the number densities of $z=7.3$ LAE, the SXDF-NB1006-2 (filled square) and 
  the SDF-NB1006-2 (open square). 
  The number densities are corrected using the detection completeness derived in \S \ref{subsec_detection_completeness}. 
  The error of the number densities includes cosmic variance and Poissonian error. The error of the Ly$\alpha$ luminosity includes 
  the spectral sky noise of both objects in the $z=7.3$ survey, and the error in the flux calibration of the DEIMOS data 
  in addition to the sky noise of SXDF-NB1006-2  (see \S \ref{sec_spectroscopy}). 
  The filled circle is the LAE at $z=6.96$ identified spectroscopically \citep{2008ApJ...677...12O}. 
  The dashed blue and red curves represent the Ly$\alpha$ LF fitted by Schechter function of LAEs at $z=5.7$ and $6.6$ 
  in the SDF, respectively \citep{2011ApJ...734..119K}. The solid blue and red curves are those at $z=5.7$ and $6.6$ 
  in the full SXDF, which is $\sim 5$ wider than the SDF \citep{2010ApJ...723..869O}. 
  The dark-green triangles, bold- and dashed-curves are photometric $z=7$ LAE samples in \citet{2010ApJ...722..803O},  
  fitted Schechter functions in the D33 and D41 field of the CFHTLS in \cite{2012ApJ...744...89H} derived by using 
  the Suprime-Cam equipped with NB973. 
  The orange filled squares, filled circles, and curve are photometric $z=7.7$ LAE samples derived by using 
  the Mayall 4m telescope/NEWFIRM camera \citep{2012ApJ...745..122K, 2010ApJ...721.1853T, 2010A&A...515A..97H}. 
  The slope parameter $\alpha$ of all the fitted Schechter functions are fixed to be $-1.5$. If SXDF-NB1006-1 
  is a real LAE at $z=7.3$ with brighter $L^{Ly\alpha}$ than one of the SXDF-NB1006-2, the cumulative number densities of 
  SXDF-NB1006-2 and SDF-NB1006-1 become the small filled square and the open square without $L^{Ly\alpha}$ error 
  slightly shifted in the Ly$\alpha$ luminosity direction for clarification.}
\label{fig_cumulative_lf}
\end{figure*}

\begin{figure*}[t]
\epsscale{0.6}
\plotone{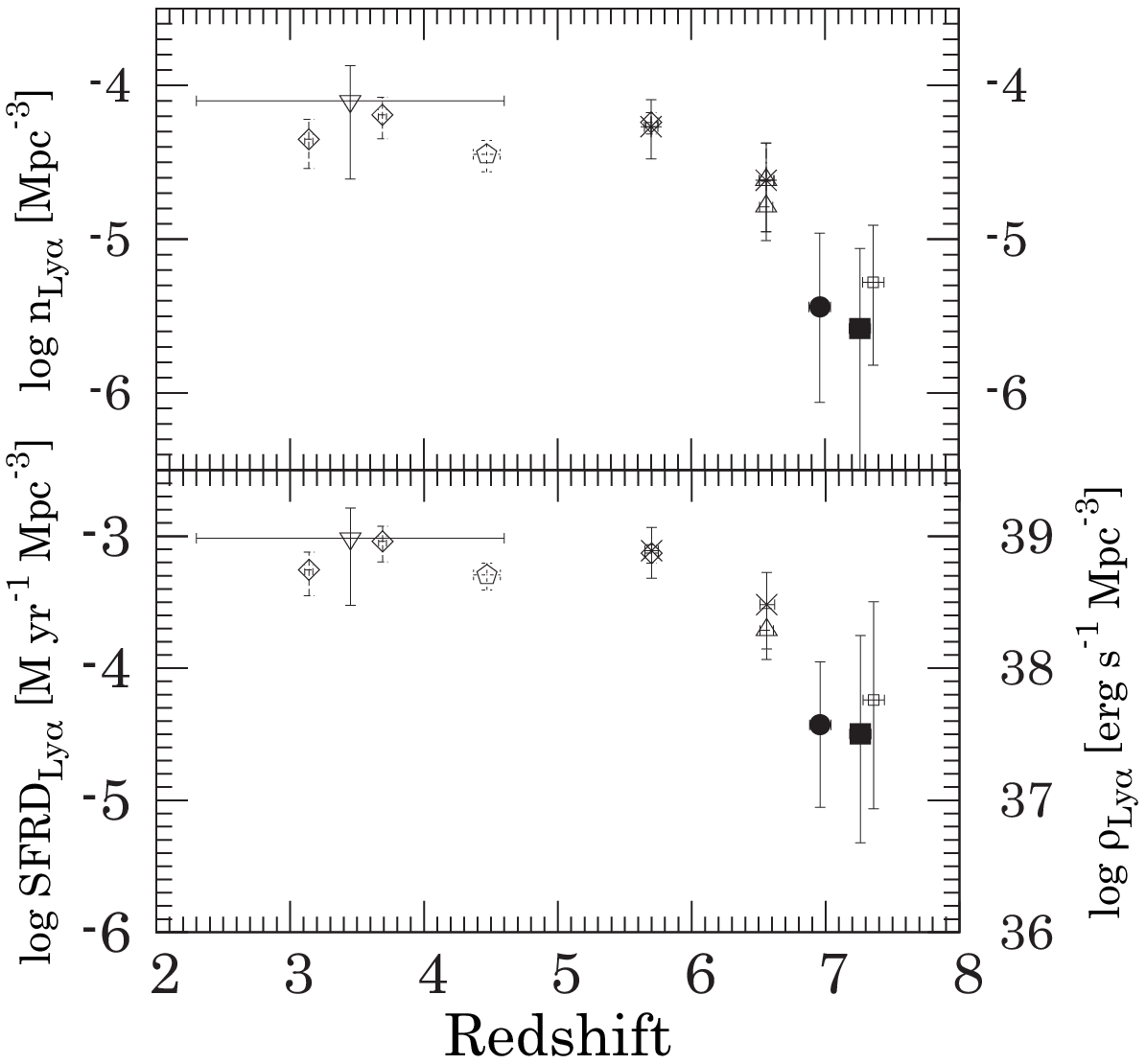}
\caption{Number density $n_{Ly\alpha}$, and star formation rate density SFRD$_{Ly\alpha}$ of the LAE at $z=7.215$, 
   SXDF-NB1006-2 confirmed spectroscopically in this study (filled square) and those at $2.3 \le z \le 7$ from the 
   spectroscopic studies down to $L^{Ly\alpha} = 1.0 \times 10^{43}$ erg s$^{-1}$. The filled circles are those of the LAE at $z=6.96$ 
   \citep{2008ApJ...677...12O}. The crosses show the densities at $z=5.7$ and $6.6$ in the SDF calculated 
   from the samples of \citep{2011ApJ...734..119K}. 
   Densities at $2.3 \le z \le 4.5$ and $z \sim 4.5$ are calculated using the best-fit Ly$\alpha$ Schechter LFs from 
   \cite{2005MNRAS.359..895V} (inverse triangles) and \cite{2007ApJ...671.1227D} (pentagons). 
   Also, densities at $z=3.1, 3.7, 5.7$ (diamonds) and $6.6$ (triangles) in the $\sim 1.0$ deg$^2$ of SXDF are calculated using the best-fit Ly$\alpha$ 
   Schechter LFs from \cite{2008ApJS..176..301O, 2010ApJ...723..869O}. Each horizontal error bar shows the redshift range of 
   each survey. The vertical error bars at $z=5.7, 6.6, 7$ and $7.3$ include both cosmic variance and Poissonian errors for 
   small number statistics. We correct for the detection completeness of the $z=7.3$ LAE survey. 
   If SXDF-NB1006-1 is a real LAE at $z=7.3$ with a brighter $L^{Ly\alpha}$ than one of SXDF-NB1006-2, 
   the number and SFR densities of SXDF-NB1006-2 becomes small the open square slightly shifted in the redshift direction 
   for clarification.}
\label{fig_density}
\end{figure*}

\subsection{The Ly$\alpha$ Luminosity Function of the $z=7.3$ LAEs}\label{subsec_lf}

In Figure \ref{fig_cumulative_lf}, we compare the luminosities of our spectroscopically confirmed LAE 
at $z=7.215$ with the previous LF results obtained at $z\le7.0$. The SDF-NB1006-2 is also included 
as a preliminary measurement. 
The number densities of both objects are corrected for the detection completeness 
derived in \S \ref{subsec_detection_completeness}. 
The error of the number density of the $z=7.3$ LAEs includes the possible
cosmic variance and the Poissonian error for small number statistics. 
The values in Table. 1 and 2 of \citet{1986ApJ...303..336G} are used as the upper and lower limit of Poissonian errors. 
The relative cosmic (root) variance $\sigma_v$ is calculated by $\sigma_v = b \sigma_{DM}$, 
where $b$ and $\sigma_{DM}$ are the bias parameter and the (root) variance of dark halo, respectively \citep{2004ApJ...600L.171S}. 
The $\sigma_{DM}$ at $z=7.3$ with our survey volume of $5.94 \times 10^5$ Mpc$^3$ is calculated to be $0.033$ by 
the model of \citet{1999MNRAS.308..119S}. The bias parameter is adopted to be $b_{max}=9$ derived from the LAE sample 
at $z=6.6$ detected in the full SXDF, which is $\sim5$ times wider than the SDF \citep{2010ApJ...723..869O}. 
The resultant cosmic variance $\sigma_v$ is $\sim 30$\%. 
The error of Ly$\alpha$ flux of the SXDF-NB1006-2 includes the error in the flux calibration from using the dwarf stars 
(see \S \ref{sec_spectroscopy}) in addition to the spectral sky noise. 
The cumulative Ly$\alpha$ LFs derived from the photometric LAE candidates at $z=7$ and $7.7$ 
\citep{2010ApJ...725..394H,2010A&A...515A..97H,2012ApJ...744...89H,2012ApJ...745..122K,2010ApJ...721.1853T}
are also plotted in Figure \ref{fig_cumulative_lf}. 
Their photometric Ly$\alpha$ LFs at $z=7$ and $7.7$ in \citet{2010A&A...515A..97H, 2010ApJ...722..803O} is based on 
Ly$\alpha$ luminosity of each LAE candidate directly converted from Ly$\alpha$ flux by assuming $f^{Ly\alpha} = f^{NB}$.
However, \citet{2010PASJ...62.1167O} estimated that $\sim77$\% and $23$\% of NB973 total flux of a $z=6.96$ LAE comes 
from Ly$\alpha$ and UV continuum fluxes, respectively, using imaging and spectroscopy data.
Hence, in Figure \ref{fig_cumulative_lf} we shifted the photometric $z=7$ and $7.7$ Ly$\alpha$ LFs faintward 
with $0.77 \times L^{Ly\alpha}$ for fair comparison with other spectroscopic Ly$\alpha$ LFs.
As shown in Figure \ref{fig_cumulative_lf}, the luminosity of SXDF-NB1006-2 is
fainter than those of the LFs at $z=5.7$ and $6.6$, though their errors are large.
Furthermore, the number density of the SXDF-NB1006-2 is slightly smaller than that of the LAE at $z=6.96$, IOK-1, 
which was selected a single pointing of NB973 imaging (only SDF) with the Suprime-Cam, 
while their Ly$\alpha$ luminosities are similar.
What needs to be emphasized is that secure LAE samples can be constructed by spectroscopy, line-asymmetry investigation, 
and checking the magnitude variability, and compared appropriately with LFs at $z\le7.0$. 
While the previous photometric LFs at $z=7$ and $7.7$ do not change significantly from $z<7$, 
the number density of the LAE at $z=7.3$ shows deficit.

\begin{figure}[t]
\epsscale{1.0}
\plotone{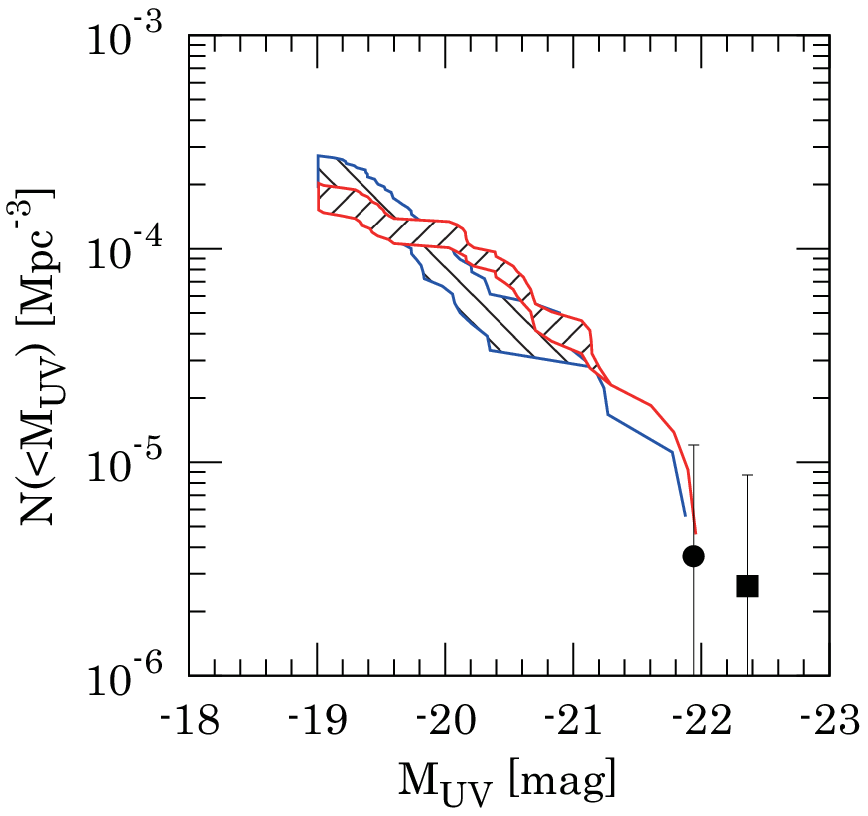}
\caption{Cumulative rest-UV continuum LFs at $z=5.7$ (blue curve) and $6.6$ (red curve) 
   derived the LAE samples in the SDF \citep{2011ApJ...734..119K}. The filled circle and square is the rest-UV number density of 
   the $z=6.96$ \citep{2008ApJ...677...12O} LAE and the SXDF-NB1006-2, respectively. 
   The UV magnitude of the SXDF-NB1006-2 is comparable with that of the rest-UV bright end at $z=5.7$ and $6.6$.}
\label{fig_uv_lf}
\end{figure}

We also compare the number and SFR densities at $z=7.3$ with those of other LAE surveys at $z\le7$ down to 
$L^{Ly\alpha} = 1.0 \times 10^{43}$ erg s$^{-1}$ in Figure \ref{fig_density} in the same way as 
\citet{2008ApJ...677...12O} to investigate the redshift-evolution of these densities. 
The densities of LAEs at $z=5.7$ and $6.6$ in \citet{2006ApJ...648....7K} are updated to those of the $\sim 80$\% 
spectroscopically identified samples \citep{2011ApJ...734..119K}, and 
the one of LAEs at $z=6.6$ selected from the $\sim5$ times wider survey area \citep{2010ApJ...723..869O} than SDF. 
It is notable that the Ly$\alpha$ LF of \citet{2010ApJ...723..869O} is mainly from the photometric LAE sample. 
The density errors are included as in Figure \ref{fig_cumulative_lf}. As shown in this figure, 
compared with number at $z=5.7$ and $6.6$, the LAE number density at $z=7.3$ decreases by factors of 
$n_{Ly\alpha}^{z=7.3}/n_{Ly\alpha}^{z=5.7} = 0.05^{+0.11}_{-0.05}$ from $z=5.7$ to $7.3$ and 
$n_{Ly\alpha}^{z=7.3}/n_{Ly\alpha}^{z=6.6} = 0.16^{+0.38}_{-0.16}$ from $z=6.6$ to $7.3$, 
although the statistical uncertainty at $z=7.3$ is large. The number and SFR density might decrease, 
$n_{Ly\alpha}^{z=7.3}/n_{Ly\alpha}^{z=7.0} = 0.72^{+0.28}_{-0.72}$ from $z=7.0$ to $7.3$.
Figure \ref{fig_density} indicates that the number and SFR densities monotonically decrease clearly from 
$z=5.7$ to $7.3$, while they do not change from $z=3$ to $6$.

In contrast, the number density of the rest-UV continuum of the SXDF-NB1006-2 at $z=7.3$ might be comparable  
with the bright end of the rest-UV continuum cumulative LFs at $z=5.7$ and $6.6$ as shown in Figure \ref{fig_uv_lf}. 
These UV continuum LFs were constructed from LAE samples in SDF \citep{2011ApJ...734..119K}. 
The UV magnitude of SXDF-NB1006-2 was derived by subtracting the Ly$\alpha$ flux from the total flux in the NB1006 magnitude.
This Ly$\alpha$ deficit might be caused by the gradual and continuous reionizing process from $z=5.7$ to $7.3$, 
because the density of the rest-UV continuum is insensitive to the surrounding HI clouds and might not evolve. 
Note however that the UV magnitude of SXDF-NB1006-2 has a large error caused by Ly$\alpha$ flux error 
(see Table \ref{table_spectroscopy}), and the Ly$\alpha$ deficit might arise from the galaxy evolution as well as HI clouds. 
This tendency is consistent with the results of the previous surveys for LAEs $z\le7$ 
\citep{2008ApJ...677...12O, 2006ApJ...648....7K, 2011ApJ...734..119K, 2010ApJ...723..869O}
and our results might show for the first time that the trend continues up to  $z\sim7.3$.
On the contrary, our result is inconsistent with the results of 
\citet{2010ApJ...725..394H, 2010ApJ...721.1853T, 2012ApJ...745..122K, 2010A&A...515A..97H} 
which reported that Ly$\alpha$ LFs do not evolve from $z=5.7$ to $7.7$. 
If the decline of Ly$\alpha$ luminosity is caused only by IGM attenuation, 
the neutral fraction $x_{HI}$ of hydrogen gas at $z=7.3$ can be estimated to be $\sim 0.79$ 
using a radiative transfer simulation based on \citet{2007MNRAS.381...75M} which calculate the number density of LAE at 
the epoch of reionization, and we assume that the IGM at $z=5.7$ is 
completely ionized. 
Compared with $L^{Ly\alpha}$ of SXDF-NB1006-2 and the Ly$\alpha$ LF at $z=7.3$ constructed by the model of 
\citet{2010ApJ...708.1119K}, the IGM transmission for Ly$\alpha$ photons $T_{Ly\alpha}^{IGM}$ can be estimated. 
Here, $T_{Ly\alpha}^{IGM}$ is calculated with $T_{Ly\alpha}^{IGM} =$ observed $L^{Ly\alpha}$ / model $L^{Ly\alpha}$. 
As we use the faintest $L^{Ly\alpha}$ in the lowest number density within their errors as the observed $L^{Ly\alpha}$, 
the derived $T_{Ly\alpha}^{IGM}$ is given as an lower limit, $0.27$. Thus, the possible $T_{Ly\alpha}^{IGM}$ ranges 
$0.27 \le T_{Ly\alpha}^{IGM} \le 1$. 
If we apply the calculated $T_{Ly\alpha}^{IGM}$ to the Ly$\alpha$ flux attenuation model of \citep{2004MNRAS.349.1137S}, 
the neutral fraction $x_{HI}$ at $z=7.3$ is estimated to be $0.82$. 
Note that the derived neutral fraction is strongly model dependent. 
However, it must be noticed that one remaining candidate, the SXDF-NB1006-1, has yet not been observed 
spectroscopically. If this candidate was a real LAE with similar Ly$\alpha$ luminosity with the SXDF-NB1006-2 at $=7.3$, 
the number density would increase to that at $z=7$, and not evolve between $z=7$ and $7.3$ as indicated by 
small points in Figure \ref{fig_cumulative_lf} and \ref{fig_density}. On the other hand, if all the candidates 
were variable objects, which indicates that no LAE at $z=7.3$ were detected in our survey volume,
the derived $x_{HI}$ is only given as an upper limit.

\section{SUMMARY and CONCLUSIONs}\label{sec_conclusion}

We have carried out a systematic imaging and spectroscopic survey for LAEs at $z \sim 7.3$, 
for the first time, in the SDF and the SXDF, using the Suprime-Cam equipped 
with new red-sensitive CCDs and a newly developed NB1006 filter. The total survey area and volume are 
$1719$ arcmin$^2$ and $5.94 \times 10^5$ Mpc$^3$, respectively. 
Four $z\sim 7.3$ LAE candidates were photometrically identified from the NB1006 imaging down to the NB 
limiting magnitude of $24.6-24.8$, and we obtained spectroscopy for three of them, SDF-NB1006-1, 2 and SXDF-NB1006-2
using Subaru/FOCAS and Keck/DEIMOS. We have detected a definitively asymmetric emission line from SXDF-NB1006-2.  
If this line is Ly$\alpha$, it is a LAE at $z=7.215$. 
SXDF-NB1006-2 has a luminosity of $1.2^{+1.5}_{-0.6} \times 10^{43} $ erg s$^{-1}$, 
a star formation rate $11$ M$_\odot$ yr$^{-1}$ estimated from the Ly$\alpha$ emission line, and a weighted skewness 
S$_\omega = 4.90\pm 0.86$. 
 
 The conclusions of this survey are summarized below. 
 
 \begin{itemize}
  \item Compared with the number density and SFR density of LAEs at $z\le7$, The number density and SFR density of 
  	LAEs monotonically decreases from $z=5.7$ to $7.3$ while they do not change from $z=3$ to $6$, and the number
	density of rest-UV continuum at $z=7.3$ is 
	comparable with the rest-UV cumulative LFs at $z=5.7$ and $6.6$. The derived decreases of the number density 
	 are $n_{Ly\alpha}^{z=7.3}/n_{Ly\alpha}^{z=5.7} = 0.05^{+0.11}_{-0.05}$ from $z=5.7$ to $7.3$ and
	 $n_{Ly\alpha}^{z=7.3}/n_{Ly\alpha}^{z=6.6} = 0.16^{+0.38}_{-0.16}$ from $z=6.6$ to $7.3$ down to $L^{Ly\alpha} = 1.0 \times 10^{43}$ erg s$^{-1}$. 
	This trend is consistent with the results of the previous surveys for LAEs at $z\le7$, and it might be caused by IGM attenuation. 
 \item The magnitude of SDF-NB1006-2 in the NB1006 images varies and it might be a AGN/QSO at $z=7.3$ or $1.7$ rather than a LAE, 
	in spite of the detected asymmetric emission line (S$_\omega = 4.59\pm 1.25$). 
	The third object, SDF-NB1006-1, is likely a galaxy with temporal luminosity enhancement associated with 
	a QSO, an AGN, or a supernova explosion, as the brightness of this object varies between the observed epochs. 
	Its spectrum does not show any emission lines. The magnitude of the candidates in the SXDF, SXDF-NB1006-1 and 2,
	did not show any significant variation, though the time span to check the variability in SXDF was not long enough 
	to completely trace the typical variability. 
 	Hence, it is important to the check whether the candidates are transient objects or not by dividing photometric images 
	into several epochs. 

 \end{itemize}
 
 In order to accurately measure the neutral fraction and reveal the relationship of LAEs and LBGs in the evolution of galaxy at 
 $z \ge 7$ in more detail,  multi-epoch surveys using next generation facilities such as the Subaru/Hyper Suprime-Cam (HSC) 
 that covers $1.5$ deg FoV in diameter,  equipped narrow band filters, are needed to establish larger LAE sample.


\acknowledgments

This paper is based on data collected with the Subaru Telescope, which is operated
by the National Astronomical Observatory of Japan. 
The analysis pipeline used to reduce the DEIMOS data was developed at UC Berkeley 
with support from NSF grant AST-0071048. This work is based in part on data obtained as 
part of the UKIRT Infrared Deep Sky Survey.
We would like to thank Yousuke Utsumi for giving us helpful advices to reduce DEIMOS spectra, 
and Tomoki Morokuma for giving us favorable comments on the AGN/QSO variability. 
We would also like to thank Caryl Gronwall for checking expressions in this paper. 
We are grateful to the Subaru Observatory staff for their help with the observations. 
T. S. acknowledges fellowship support from the Japan Society for the Promotion of Science. 
This research was supported by the Japan Society for the Promotion of Science through 
Grant-in-Aid for Scientific Research 23340050.

{\it Facilities:} \facility{Subaru/Suprime-Cam (NAOJ)}, \facility{Subaru/FOCAS (NAOJ)}, 
\facility{Keck/DEIMOS}.


\bibliographystyle{bst/apj}
\bibliography{reference}

\end{document}